\documentclass[preprint2]{aastex}

\shorttitle{Heating the Solar Atmosphere}
\shortauthors{Dumin}

\begin{document}

\title{Heating the Solar Atmosphere\\
by the Self-Enhanced Thermal Waves\\
Caused by the Dynamo Processes}

\author{Yurii V. Dumin}
\affil{Institute of Terrestrial Magnetism, Ionosphere, and
Radio Wave Propagation (IZMIRAN),\\
Russian Academy of Sciences,\\
Troitsk, Moscow reg., 142190 Russia}

\email{dumin@yahoo.com, dumin@izmiran.ru}

\begin{abstract}
We discuss a possible mechanism for heating the solar atmosphere
by the ensemble of thermal waves, generated by the photospheric dynamo
and propagating upwards with increasing magnitudes. These waves are
self-sustained and amplified due to the specific dependence of
the efficiency of heat release by Ohmic dissipation on the ratio of
the collisional to gyro- frequencies, which in its turn is determined
by the temperature profile formed in the wave. In the case of
sufficiently strong driving, such a mechanism can increase the plasma
temperature by a few times, i.e.\ it may be responsible for heating
the chromosphere and the base of the transition region.
\end{abstract}


\section{INTRODUCTION}

Seeking for the mechanism of heating the solar atmosphere with height
from a few thousand to a million Kelvin is one of long-standing
problems in astrophysics. The approaches proposed by now, roughly
speaking, can be separated into two groups
\citep[e.g.\ reviews by][and references therein]{wal03,erd07}.
The first group deals with a generation of some kinds of
(magneto-) hydrodynamic waves or pulses in the base of the solar
atmosphere and their subsequent propagation and dissipation in
the upper layers.
The mechanisms of the second group assume that the heating is
due to the ensemble of small-scale flare-like events, caused by
the reconnection processes in the specific magnetic field configurations.
It is commonly believed now that no unique mechanism can provide
the entire heating of the solar atmosphere, and a few of them are
acting simultaneously in the Sun.

Besides, as was proposed by \citet{asc07}, the inverted (increasing
with height) temperature profile might be formed ``dynamically''---due
to the fluxes of plasma heated in the lower layers and propagating
upwards (which is often called evaporation). This is, in fact,
the third kind of the heating mechanisms.

The aim of the present work is to consider one more mechanism, which,
strictly speaking, belongs to none of the three above-mentioned
groups but inherits some features from all of them.
Like mechanisms from group~II, it assumes an ensemble of
the small-scale flare-like events; but such events require no
special field topology for their occurrence (as in the case of
reconnection) and are more similar to the specific type of
wave pulses, as in the mechanisms of group~I. At last, our
scenario reminds the Aschwanden's approach (group~III),
because the heating process propagates dynamically from
the lower to upper layers.

\section{FORMULATION OF THE MODEL}

\begin{figure*}[t]
\epsscale{1.75}
\plotone{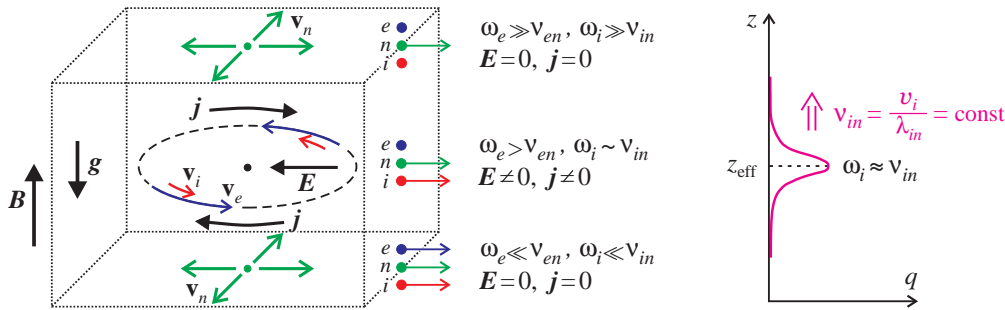}
\caption{Sketch of the dynamo processes in the partially-ionized
plasma (motion of the neutrals, electrons, and ions are colored
green, blue, and red, respectively, in the online version of this
figure) and a schematic structure of the associated thermal wave,
propagating upwards (colored magenta in the online version).
\label{fig:dynamo_sketch}}
\end{figure*}

The dynamo processes in weakly-ionized plasmas, considered in this
section, closely remind the so-called $ S_q $-dynamo, well known
in the ionospheric physics
\citep[e.g.\ review by][and references therein]{ric89}.
Similar processes in the solar physics are called photospheric
dynamo, and they were employed in a number of papers to describe
generation of field-aligned currents and perturbations of
the magnetic fields \citep{hen87}. The associated thermal
effects were usually neglected; an exception is the old work
by \citet{sen72}.

\subsection{Qualitative Description}

Let us consider a column of partially-ionized plasma stratified by
the gravitational field~$ {\bf g} $ and permeated by the magnetic
field~$ {\bf B} $, which for simplicity is taken to be vertical
and constant. Besides, we assume that this plasma is driven radially
in the horizontal plane by the motion of a neutral
component~$ {\bf v}_n $ (which can be associated, for example,
with convective flows at the photospheric level);
see Figure~\ref{fig:dynamo_sketch}.

Due to the gravitational stratification, a neutral gas density
sharply decreases with height. Respectively, the free-path length
and the time of interparticle collisions sharply increase, while
the collisional frequency decreases. Then, at the bottom of
the column, where the frequencies of collisions of both electrons
and ions with neutrals, $ {\nu}_{en} $ and $ {\nu}_{in} $ respectively,
are much greater than their gyrofrequencies~$ {\omega}_e $
and~$ {\omega}_i $, the charged particles are completely dragged
by the neutrals and move with them; so that no local electric
fields and currents are generated.

In the opposite case, at the top of the column, where
the collisional frequencies are much less than the gyrofrequencies,
the electrons and ions are unaffected by the radial motion of
the neutral component and, therefore, will remain at rest.
No electric fields and currents are produced again.

The most nontrivial situation arises at the intermediate heights, where
$ {\nu}_{in} \! \sim \! {\omega}_i $ and $ {\nu}_{en} \! < \! {\omega}_e $.
Under these circumstances, the ions are dragged by neutrals,
while the electrons remain approximately at rest.
Due to the separation of the electrons and ions, a radial
electric field~$ \bf E $ should develop, resulting in its turn in
the azimuthal drift of the charged particles. However, while
the electrons will experience such a drift almost without
resistance, the ions will be decelerated by the collisions with
neutrals. Therefore, the azimuthal (Hall) electric currents~$ \bf j $
are generated, as shown in Figure~\ref{fig:dynamo_sketch}.

The above-mentioned electric fields and currents are concentrated in
a quite narrow height range, where $ {\nu}_{in} \! \sim \! {\omega}_i $;
and they should lead, firstly, to the magnetic-field perturbations
\citep[considered by][]{hen87} and, secondly, to the dissipative
effects and heat release, which are just the subject of our study.
(The dissipation exists due to the azimuthal electric field appearing
in the coordinate frame co-moving with plasma.)

Once a heat release has occurred at the height where
$ {\nu}_{in} \! \approx \! {\omega}_i $ (we shall call it~$ z_{\rm eff} $),
this spot is no longer efficient for the subsequent heat release,
because the plasma is heated and the thermal velocity of its
particles~$ v_i $ as well as the collisional frequency
$ {\nu}_{in} \! = v_i / {\lambda}_{in} $
increased (under assumption of time-independent neutral density~$ n_n $,
constant ion-neutral cross-section~$ {\sigma}_{in} $ and, consequently,
the unchanged mean free path
$ {\lambda}_{in} \! = 1 / ( {\sigma}_{in} n_n ) $).
Therefore, the condition $ {\nu}_{in} \! \approx \! {\omega}_i $ will
no longer be satisfied.

On the other hand, due to the heat conduction in space,
the ion thermal velocity~$ v_i $ will increase also at a greater
height (where the mean free path~$ {\lambda}_{in} $ is larger due to
the initial gas stratification) and, therefore, the spot of effective
heat release,
$ {\nu}_{in} \equiv v_i / {\lambda}_{in} \approx {\omega}_i $,
will move there. As a result, we should expect formation of the specific
self-sustained thermal wave propagating upwards, as depicted in
the right-hand panel of Figure~\ref{fig:dynamo_sketch}. Moreover,
the wave amplitude will increase with height, just because of decreasing
a heat capacity per unit volume in the stratified gas.

It should be also mentioned that heat release in the dynamo-region
may be further facilitated by the onset of two-stream instability
\citep[][and references therein]{sen72}; but we shall not consider
here in more detail the corresponding microphysical mechanisms.

Finally, let us discuss more carefully the scope of applicability
of our scenario. In fact, the ion's collisional frequency is composed
of two parts:
$ {\nu}_i \! = {\nu}_{in} \! + {\nu}_{ii} $,
where $ {\nu}_{in} $ and $ {\nu}_{ii} $ are the frequencies of
ion-neutral an ion-ion collisions. It was assumed everywhere above
that the second term can be ignored. On the other hand, the degree of
ionization increases with height; so that the ion-ion collisions will
inevitably become dominant starting from some altitude. In this case,
$ {\lambda}_{ii} \! = 1 / ( {\sigma}_{ii} n_i ) \sim  v_i^4 / n_i \, $,
since~$ {\sigma}_{ii} \! \sim v_i^{-4} $. So, the ion's collisional
frequency
$ {\nu}_{ii} \! = v_i / {\lambda}_{ii} \sim n_i / v_i^3 $
decreases with temperature, as distinct from the case of ion-neutral
collisions. However, this fact is unrelated to the criterion of
the dynamo-layer development, $ {\nu}_{in} \! \sim \! {\omega}_i $,
because collisions between the charged particles of the same kind
cannot affect the generation of electric currents. The dominant role
of ion-ion collisions can change only the coefficient of heat
conductivity.

Therefore, the scope of applicability of the presented model is
somewhat wider than seems at the first sight. At the same time,
the plasma must be weakly ionized in the entire height range
under consideration, since otherwise it would be meaningless
to speak about dragging the charged particles by the neutral gas.

\subsection{Basic Equations}

In the simplest one-dimensional approximation, the process of
heat transfer along the magnetic flux tube can be described by
the equation:
\begin{equation}
\frac{\partial}{\partial t} \Big[ c \, n \, T \Big] -
\frac{\partial}{\partial z}
\Big[ c \, n \, \chi \, \frac{\partial}{\partial z} \, T \Big] \! = \, q \, ,
\label{eq:heat_transfer}
\end{equation}
where
$ T $~is the temperature of the heavy particles, i.e.\ neutrals and ions
(the electron temperature is not of interest here),
$ t $~is the time,
$ z $~is the vertical coordinate,
$ n $~is the number density of the heavy particles,
$ \chi \! \approx \! (1/3) \lambda v $~is the temperature conductivity,
$ c \! \equiv \! {\rm const} $~is the heat capacity per a heavy particle,
and
$ q $~is the volume density of the heat release due to Ohmic dissipation.
(The basic parameters of neutrals and ions are assumed to be
approximately the same and, therefore, written without additional
subscripts.)

Leaving aside a microscopic theory of the Ohmic heating, let us use
a phenomenological approximation of the heat-release profile,
based on its qualitative behavior discussed above. Namely, we take
Taylor expansion in the vicinity of the heat-release maximum,
achieved at
$ {\nu}_{in} \! \approx {\omega}_i $:
\begin{equation}
q (z, T) = q_0 \bigg\{ 1 \! - \alpha {\bigg( 1 -
  \frac{{\nu}_{in} (z, T)}{{\omega}_i} \bigg)}^{\!\! 2} \! +
  \, \dots \bigg\} \, ,
\label{eq:heat_release}
\end{equation}
where $ q_0 $~is the amplitude of the heat release, and
$ \alpha $~is the characteristic width of the profile.

Next, we shall use for simplicity the time-independent exponential
(i.e., actually, isothermal) height profile of the gas density:
\begin{equation}
n(z) = n_0 \, \exp \! \big[ \! - \! z / H \big] \, ,
\label{eq:heigh_profile}
\end{equation}
where $ H = T_0 / mg \equiv {\rm const}$ is the height scale.
(In other words, it is assumed that the initial gas profile is not
changed during the heat propagation; of course, redistribution of
the gas density should be taken into account in a more accurate
treatment.)

The thermal velocity is evidently related to the temperature as
\begin{equation}
v = v_0 \sqrt{ T / T_0 } \, .
\label{eq:therm_vel}
\end{equation}
So, assuming that the cross
section~$ \sigma $ is constant (which is a reasonable approximation
for the collisions of ions with neutrals and neutrals with each other),
we get:
\begin{equation}
{\nu}_{in} = \sigma n v_0 \sqrt{ T / T_0 } =
{\nu}_{in0} ( n / n_0 ) \sqrt{ T / T_0 } \, .
\label{eq:coll_freq}
\end{equation}
Besides, it is convenient to choose the origin of $z$-axis
($ z \! = \! 0 $) in the point where $ {\nu}_{in0} \! = \! {\omega}_i $.

At last, substituting
expressions~(\ref{eq:heat_release})--(\ref{eq:coll_freq}) into
equation~(\ref{eq:heat_transfer}), we obtain the equation of
heat transfer in the following form:
\begin{eqnarray}
e^{- \! z^* \! / \! H^*} \, \frac{\partial T^*}{\partial t^*} -
\frac{1}{3} \, \frac{\partial}{\partial z^*}
\Big[ T^{* 1/2} \, \frac{\partial T^*}{\partial z^*} \Big] \, = \quad
\nonumber
\\[1.5ex]
q_0^* \Big\{ 1 - \alpha \big( 1 - e^{- \! z^* \! / \! H^*} \,
T^{* 1/2} \big)^2 + \dots \Big\} \, ,
\label{eq:heat_transfer_dimless}
\end{eqnarray}
where the dimensionless quantities (marked by asterisks) were
introduced as
\begin{equation}
T^* \! = T / T_0 \, , \quad
t^* \! = t / {\tau}_0 \, , \quad
z^* \! = z / {\lambda}_0 \, ,
\label{eq:definitions_1}
\end{equation}
\begin{equation}
H^* \! = H / {\lambda}_0 \, , \quad
q_0^* = \frac{ q_0 {\tau}_0 }{ c \, n_0 T_0 } \, ,
\label{eq:definitions_2}
\end{equation}
$ T_0 $~is the initial temperature of the gas,
$ {\tau}_0 \! \equiv {\nu}_{in0}^{-1} $, and
$ {\lambda}_0 \! \equiv {\lambda}_{in0} \, $.

\section{RESULTS OF THE COMPUTATION}

Expression~(\ref{eq:heat_transfer_dimless}) represents a quite specific
type of the nonlinear parabolic partial differential equation.
A convenient tool for solving just this type of problems was
implemented in the NAG numerical library---this is the subroutine
D03PCF \citep{nag01,ber90}.

We restrict our consideration here by the simplest numerical
solution, just to illustrate how the proposed mechanism works.
(A detailed treatment of more realistic situations
will be published elsewhere.) So, let us take the initial
condition in the form:
\begin{equation}
T^* = 1 \quad {\rm at} \quad t^* = 0
\label{eq:init_cond}
\end{equation}
(i.e.\ the entire gas had initially the same temperature~$ T_0 $).

The boundary conditions will be specified, somewhat arbitrarily,
as vanishing heat fluxes at the top and bottom of the computational
region:
\begin{equation}
\frac{\partial T^*}{\partial z^*} \! = 0 \quad
{\rm at} \quad z^* \! = z_d^* \quad {\rm and} \quad z^* \! = z_u^* \, .
\label{eq:bound_cond}
\end{equation}
(However, we shall discuss below only behaviour of the solution
far from the boundaries, which is insensitive to the boundary
conditions.)

\begin{figure}[t]
\epsscale{1.0}
\plotone{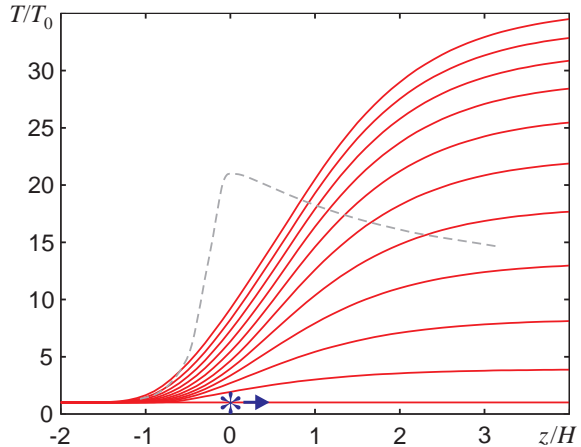}
\caption{Computed height profiles of the temperature at the successive
instants of time (colored red in the online version of the figure).
The star (colored blue) shows the spot of the initial heat release,
subsequently moving to greater heights; and the dashed (grey) curve
represents a hypothetical temperature profile that would be formed in
the case of a static heat source.
\label{fig:results_comput}}
\end{figure}

At last, let us take, for example, the following parameters
of the problem:
$ H^* \! = 10 $, $ q_0^* \! = 0.1 $, $ \alpha \! = 1 $,
$ z_d^* \! = -100 $, and $ z_u^* \! = 100 $.
The obtained numerical solution is presented in
Figure~\ref{fig:results_comput}, where height profiles of
the temperature are drawn for the successive instants of time
with increments $ \Delta t^* \! = 10 $.

As is seen, a propagation of the self-sustained thermal wave really
results in the formation of temperature profiles monotonically
increasing with height above the spot of the initial heat release
at $ z \! = 0 $. This is essentially related to the dynamical nature
of the process; otherwise, at the fixed heating source,
the temperature profile would be strongly asymmetric (because of
the substantial height dependence of the heat conductivity),
but its maximum would always take place at the fixed site of
the heat release.

Next, the series of profiles in Figure~\ref{fig:results_comput}
monotonically increases with time, without any clear evidence for
saturation. This is just because our simplified model does not take
into account the radiative loss as well as exhausting efficiency of
the dynamo processes in the course of time. It would be desirable,
of course, to include into the future numerical simulations a realistic
function of radiative loss (which should be especially important at
the enhanced degree of ionization, associated with the increased
plasma temperature). However, the role of radiative cooling should not
be overestimated in advance: when the plasma temperature is well below
the ionization thresholds of the plasma elements, then the establishment
of higher ionization density requires a very large number of
interparticle collisions and, therefore, may take a considerable time.
So, if the discussed thermal waves are impulsive phenomena, then
there may be just insufficient time for the increase in ionization,
and the radiative losses will be not so large.

\section{DISCUSSION AND CONCLUSIONS}

As follows from the above consideration, the presented model may be
a mechanism for heating the lower part of the solar atmosphere,
i.e.\ formation of the temperature profile increasing with height.
At the same time, it should be borne in mind that it works only in
weakly-ionized plasma. To provide plasma heating up to the coronal
temperatures, it is necessary to invoke other heat-release mechanisms,
commonly based on the magnetic reconnection
\citep[for the review of recent results, see][]{shi07,aul07}.

Let us mention also that some qualitative ideas why heating of
the solar plasma should occur just at the spot of approximate
equality between the collisional and gyro- frequencies were
put forward in our paper about a decade ago \citep{dum02}.
Unfortunately, it was not clear from that work how is it possible
to get the temperature profile monotonically increasing with height
rather than peaking at a fixed altitude of the maximum heat release.
The mechanism of the propagating thermal wave, described in
the present paper, gives a possible answer to that question.

\acknowledgments

I am grateful to B.V.~Somov for valuable discussions, as well as
to V.N.~Shubin and H.~Scherrer-Paulus for help in the NAG software
handling.

\end{document}